\documentclass[aps,pre,reprint,superscriptaddress,10pt]{revtex4-1}
\usepackage{amsmath}
\usepackage{amssymb}
\usepackage{graphicx}
\usepackage{dcolumn}
\usepackage{bm}
\usepackage{hyperref}
\usepackage{courier}
\usepackage{braket}
\usepackage{float}
\usepackage[usenames,dvipsnames]{xcolor}
\hypersetup{
  colorlinks   = true,	
  urlcolor     = blue,	
  linkcolor    = blue,	
  citecolor   = blue	
}
\begin{document}
\newcommand{\ve}{\varepsilon}
\title{Reverse quantum aging: Restoring rhythmicity from quantum aging state}

\author{Bulti Paul}
\affiliation{Chaos and Complex Systems Research Laboratory, Department of Physics, University of Burdwan, Burdwan 713 104, West Bengal, India}
\author{Biswabibek Bandyopadhyay}
\affiliation{Department of Physics \& Astronomy, Northwestern University, 2145 Sheridan Rd. Evanston, IL 60208 USA}
\author{Tanmoy Banerjee}
\email[]{tbanerjee@phys.buruniv.ac.in}
\thanks{he/him/his}
\affiliation{Chaos and Complex Systems Research Laboratory, Department of Physics, University of Burdwan, Burdwan 713 104, West Bengal, India}

\date{\today}

\begin{abstract}
Quantum aging transition has recently been reported, whereby the oscillatory dynamics of a network of globally coupled quantum self-sustained oscillators is degraded sharply when the fraction of inactive nodes exceeds a critical value. Here we explore the reverse process: the revival of rhythmic dynamics from an aged state in a quantum oscillator network by introducing a controlled incoherent single-boson pumping channel. Using a self-consistent quantum master equation in the  Lindblad form for globally coupled active–inactive quantum Stuart–Landau oscillators, we show that increasing the pumping strength progressively suppresses the aging transition and eventually eliminates it, restoring rhythmicity even when all oscillators are inactive. Remarkably, complete compensation of the coupling-induced dissipative loss is not required; a substantially weaker pumping strength is sufficient to eliminate the aging transition. We show that the revival is governed by a redistribution of the Fock-state population of the inactive oscillators toward higher-occupation states. Our results establish controlled incoherent pumping as an effective mechanism for restoring dynamical robustness in quantum oscillator networks.
\end{abstract}

\maketitle

\section{Introduction}
Degradation or {\it aging} of nodes often results in poor dynamical robustness of a network of coupled oscillators that may lead to system malfunction or collapse. In this context, Daido and Nakanishi \cite{daido} discovered an interesting emergent behavior, namely the aging transition where the macroscopic oscillatory state collapses to a globally cessation state due to the interplay of active-inactive oscillators. Later on, a series of studies revealed that classical aging transition is indeed a general dynamical behavior that arises from a multiple coupling functions as well as a wide class of classical oscillators \cite{daido-pre,daido-hetero} (see \cite{majhi24} and references therein).

The quantum version of the aging transition has been discovered recently by \citet{qaging} that rekindles the topic of dynamical robustness in quantum network. It was shown in \cite{qaging} that unlike classical scenario, quantum aging is not characterized by a complete cessation to a collapsed state but a rapid decrease in the mean boson number with increasing fraction of inactive oscillator. Moreover, there exists a `knee point' around which the aging transition occurs. Later, the aging transition is reported in coupled qubits as well confirming the generality of the aging phenomena in the quantum domain \cite{qubit-aging,qbit-aging-2}. Recently, coherent atom--oscillator interactions were shown to promote the quantum aging transition such that aging occurs even with a smaller fraction of inactive oscillators \cite{qbit-aging-2}. The continuous efforts of understanding the aging transition and other well-known emergent behaviors  such as quantum synchronization \cite{lee_prl,brud_prl1,qsenthil-1,bruder-nonreciprocal-PRA}, symmetry-breaking transitions \cite{qmod,qturing,qkerr,qsenthil-2,qrev,paul2024attractive}, and amplitude death \cite{qad1,qad2} in quantum domain is due to the fact that nonlinear dynamics in open quantum systems have shown several novel and counter intuitive behaviors with no direct classical analogue \cite{qsync-Review-26,chia,chia-25}. Recent experimental realizations of quantum synchronization and related nonlinear dynamics demonstrate that these phenomena are experimentally observable in physical quantum systems, establishing their relevance beyond idealized theoretical models \cite{expt-kanpur-sync,expt-ibm-sync,science-expt-vdp,expt-sync-qvdp}.

In this context, maintaining rhythmic dynamics in an oscillator network is of central interest in the study of coupled oscillators as cessation of oscillations may lead to irrecoverable system degradation \cite{kurths_pg} or pathological malfunction \cite{cardio,brain}. Therefore, controlling the aging state and reviving oscillations from an aged state have been a challenging task for the researchers for decades; to this endeavor several techniques have been proposed in classical domain where mostly the coupling between the oscillators are modified in a controlled way \cite{natcom,recovery-pre13,liu2016enhancing,kundu2018resumption,sharma2021enhancement,premraj-aging} (see \cite{dvprep} for a recent review). 

Like classical systems, controlling aging and its removal is important in quantum systems as well since practical quantum systems also suffer unwanted losses in the form of dielectric loss in superconducting junction in circuit quantum electrodynamics \cite{cQED}, lossy cavity \cite{loss-cavity}, mechanical dissipation \cite{opto-revmodphys}. This invokes the question of revoking of quantum aging transition and reestablishing the rhythmic behaviors in a quantum network.
However, in the quantum domain, to the best of our knowledge, no technique of revoking the aging transition and thus improving the robustness of an oscillator network has been reported yet. The classical techniques of revival of oscillations from the aged state are less effective and can not be extended to the quantum domain due to their limited applicability and feasibility of practical realization in quantum systems.

In this paper, we propose a technique for the revival of rhythmic dynamics from the quantum aging state by controlled incoherent single-boson pumping that is of quantum origin and is suitable for practical realization. We consider a network of active-inactive quantum Stuart-Landau oscillators under global diffusive coupling and using the formalism of open quantum systems demonstrate that quantum aging state can be revoked in a controlled manner to revive rhythmicity by controlled interaction of the network of quantum nodes with the environment through incoherent single-boson pumping. 
Significantly, the revival occurs at a pumping strength substantially below the full-compensation limit, highlighting the efficiency of the proposed control mechanism. Apart from characterizing the revival across the relevant parameter space, we identify its microscopic origin in the redistribution of the Fock-state population of the inactive oscillators, providing a physical picture of how environmental engineering can reverse the aging process. The proposed mechanism therefore provides a controllable route toward restoring collective rhythmicity in experimentally accessible quantum oscillator networks.

\section{Mathematical model: Quantum master equation}
The quantum master equation of $N$ diffusively coupled quantum Stuart-Landau oscillators under diffusive global coupling and the proposed single boson incoherent pumping is given by,
\begin{multline}
\begin{split}
\label{eq:qm}
\dot \rho &= \sum_{j=1}^{N} \Bigl(-i[ \omega a_j^\dagger a_j,\rho] + G_j \mathcal{D} [O_j](\rho) + k \mathcal{D} [{a_j}^2] (\rho)\Bigl)\\
&+ \frac{V}{N} \sum_{j=1}^{N} \sideset{}{'} \sum_{{j^\prime}=1}^{N} \mathcal{D} [a_j - a_{j^\prime}](\rho)
+\underbrace{\Gamma_{\rm p} \sum_{j=1}^{N} \mathcal{D} [a_j^\dagger] (\rho)}_{\text{Control term}}.
\end{split}
\end{multline}
Here $ a_j $ and $ a_j^\dagger $ are the bosonic annihilation and creation operator, respectively  of the $j$-th oscillator. $\mathcal{D} [\hat{L}]$ is the Lindblad dissipator having the form $ \mathcal{D} [\hat{L}] (\rho) = \hat{L} \rho \hat{L^\dagger} - \frac{1}{2} \{\hat{L^\dagger}\hat{L},\rho\} $, where $ \hat{L} $ is an operator. The operator $ O_j $ is introduced to bring the notion of active-inactive element (will be elaborated later). $ k $ is the nonlinear damping rate governing the two-boson loss. The second term represents the diffusive global dissipative coupling with coupling strength $V$. Here the prime in one of the summations indicates that the sum does not contain $ j^\prime = j $. 

We introduce the last term (shown in the under brace) that represents the single boson (photon, phonon, or other bosons) pumping in a controlled rate that is determined by the factor $\Gamma_{\rm p}$: note that the dissipator $\mathcal{D} [a_j^\dagger]$ represents a single bosonic channel through which the system and the environment interacts. For $\Gamma_{\rm p}=0$, Eq.~\ref{eq:qm} gives quantum Stuart-Landau oscillators under diffusive global coupling without any additional boson pumping, which was studied in Ref.~\cite{qaging} in the context of quantum aging. In this paper our main concern is to explore the effect of the rate of single boson pumping $\Gamma_{\rm p}$ in controlling and revoking the aging phenomenon.

The notion of active and inactive element is determined by the operator of the Lindblad dissipator associated with the coefficient $G_j$ in the following manner
$$
O_j = \left\{
    \begin{array}{ll}
        a_j^\dagger & \mbox{for active oscillators} \\
        a_j & \mbox{for inactive oscillators}
    \end{array}
\right.
$$
The network consists of two groups: $ N_a $ number of nodes are active having rate $ G_j = G_a$ ($j \in \{1, 2, ..., N_a\}$) and $ N_i $ number of nodes are inactive having rate $ G_j = G_b $ ($j \in \{N_a+1, ..., N\}$); thus $ N_a + N_i = N $. The fraction of inactive nodes in the network is given by $p = \frac{N_i}{N} = \frac{N-N_a}{N}$. Therefore, $0\le p \le 1$: $p=1$ means all the nodes are {\it inactive} whereas $p=0$ indicates the whole network consists of {\it active} nodes.

For a large number of identical nodes, we can factorize the density matrix of the many-body system as $\rho\approx\otimes_{j=1}^{N}\rho_j$. This corresponds to the mean-field approximation where each oscillator interacting with the mean-field and the reduced master equation (\ref{eq:qm}) reads (see Appendix~\ref{app:a}) 
\begin{multline}
\begin{split}
\label{eq:3}
\dot \rho_j &= -i[ \omega a_j^\dagger a_j,\rho_j] + G_j \mathcal{D} [O_j](\rho_j) + k \mathcal{D} [{a_j}^2] (\rho_j)\\
&+ \frac{2V(N-1)}{N} \mathcal{D} [a_j] (\rho_j) + V (\mathcal{A}[a_j^\dagger,\rho_j] - \mathcal{A}^* [a_j, \rho_j] )\\
&+ \Gamma_{\rm p} \mathcal{D} [a_j^\dagger] (\rho_j),
\end{split}
\end{multline}
where $ \mathcal{A} $ and $ \mathcal{A}^* $ have the forms: $ \mathcal{A} = \frac{1}{N} \sideset{}{'} \sum_{{j^\prime}=1}^{N} {\langle a_{j^\prime} \rangle}_j $ and $ \mathcal{A^*} = \frac{1}{N} \sideset{}{'} \sum_{{j^\prime}=1}^{N} {\langle a_{j^\prime}^\dagger \rangle}_j $. Here $ \langle \hat{L} \rangle_j $ denotes the average of operator $ \hat{L} $ with respect to the one-body density matrix $ \rho_j $.

We parameterize the controlled single-boson pumping rate as
\begin{equation}
\Gamma_{\rm p}=\frac{2V(N-1)}{N}\gamma,
\qquad 0\leq\gamma\leq1,
\end{equation}
where $\gamma$ denotes the fraction of the coupling-induced dissipative
rate compensated by the additional incoherent pumping channel. This
normalization is natural because the same rate
$2V(N-1)/N$ ($=\kappa_c$) already appears as the local dissipative contribution
generated by the global diffusive coupling in Eq.~(\ref{eq:3}): therefore, $\gamma$ measures the pumping strength relative to the coupling-induced dissipative scale. Thus $\gamma=0$
corresponds to the uncontrolled network, whereas $\gamma=1$ corresponds to complete compensation of this coupling-induced linear damping. 
In the next we will show that in order to revoke aging and revive rhythmicity, remarkably the bosonic pumping introduced needs not to fully compensate the coupling-induced linear damping but rhythmicity is established for much weaker bosonic pump.

\section{Linear stability of the coherent amplitude dynamics}
To obtain an analytical prediction for the destruction of the aging state, and revival of rhythmicity we analyze the stability of the homogeneous zero-amplitude state of the
active--inactive quantum network described by Eq.~(\ref{eq:3}). We introduce the coherent amplitudes of the two populations as $\alpha_s=\langle a\rangle_s={\rm Tr}(a\rho_s)$ ($s\equiv a,i$): $\alpha_a=\langle a\rangle_a$ (for active nodes), and $\alpha_i=\langle a\rangle_i$ (for inactive nodes). Within the mean-field description, the self-consistent fields entering Eq.~(\ref{eq:3}) are
\begin{align}
\mathcal{A}_a &=\left(1-p-\frac{1}{N}\right)\alpha_a+p\alpha_i,\label{eq:Aa}
\\
\mathcal{A}_i &=(1-p)\alpha_a+\left(p-\frac{1}{N}\right)\alpha_i.\label{eq:Ai}
\end{align}
The nonlinear dissipator $k\mathcal{D}[a^2]$ gives a contribution
proportional to $\langle a^\dagger a^2\rangle$ in the equation for the
first moment and therefore, does not contribute to the linearized
dynamics about the zero-amplitude state. Retaining only terms linear in
$\alpha_a$ and $\alpha_i$, the equations for the active and inactive population become
\begin{equation}
\begin{aligned}
\dot{\alpha}_a &=\left(\frac{G_a}{2}+V\left(r\gamma-p\right)-i\omega\right)\alpha_a+Vp\alpha_i,
\\
\dot{\alpha}_i &=V(1-p)\alpha_a+\left(-\frac{G_b}{2}+V\left(p-1+r\gamma\right)-i\omega
\right)\alpha_i .
\end{aligned}
\label{eq:active_inactive}
\end{equation}
Here $r=\frac{N-1}{N}$. Equations~\eqref{eq:active_inactive} can be expressed compactly as $\dot{\bm{\alpha}}=\bm{M}\bm{\alpha}$, 
$\bm{\alpha} = \begin{pmatrix} \alpha_a & \alpha_i \end{pmatrix}^\intercal$
with the stability matrix $\bm{M}=-i\omega \bm{I}+\begin{pmatrix}
M_0 & Vp\\
V(1-p) & M_1
\end{pmatrix},$
where $M_0=\frac{G_a}{2}+V\left(r\gamma-p\right)$, and $M_1=-\frac{G_b}{2}
+V\left(p-1+r\gamma\right)$. The eigenvalues of $\bm{M}$ are $\lambda_{\pm}=-i\omega+\frac{M_0+M_1}{2}\pm\frac{1}{2}\sqrt{(M_0-M_1)^2+4V^2p(1-p)}$. Note that the frequency term $-i\omega$ only shifts the imaginary part of the eigenvalues and therefore does not affect the linear stability. The stability boundary is consequently determined from the real matrix in $\bm{M}$: At the onset of instability, the largest real eigenvalue vanishes. An eigenvalue analysis gives the critical inactive fraction $p_c^{\rm lin}$; In the thermodynamic limit, $N\rightarrow\infty$, so that $r\rightarrow1$, we get $p_c^{\rm lin}$ 
\begin{equation}
p_c^{\rm lin}=\frac{\left(G_a+2V\gamma\right)\left(G_b+2V(1-\gamma)\right)}{2V(G_a+G_b)}.
\label{eq:pc_thermo}
\end{equation}

Now we define the critical threshold value of the control parameter $\gamma_c^{\rm lin}$ such that it can revive the network from the completely aged state: this happens when the critical inactive fraction reaches the limiting value $p_c^{\rm lin}=1$. From Eq.~\eqref{eq:pc_thermo}, this condition is
\begin{equation}
\gamma_c^{\rm lin}=\frac{G_b}{2V}.
\label{eq:gc_thermo}
\end{equation}
Equations~\eqref{eq:pc_thermo} and \eqref{eq:gc_thermo} provide an analytical
linear-stability description of the control-induced revocation of the aging state: When $\gamma$ is increased beyond the critical value in Eq.~\eqref{eq:gc_thermo}, the inactive population becomes linearly unstable to the zero-amplitude state even in the limit $p\rightarrow1$ and thus aging state is completely revoked.

Note that this linear result is independent of $k$, since $k\mathcal D[a^2]$ contributes only at nonlinear order about the zero-amplitude state and therefore does not appear in $\gamma_c^{\rm lin}$. In the next section we will see that the actual quantum-aging boundary (denoted by $\gamma_c$) is shifted towards a higher value, i.e., $\gamma_c>\gamma_c^{\rm lin}$ since the nonlinear damping parameter $k$ plays a crucial role in quantum aging and its revival.

\section{Analysis and results in the Quantum domain}

To quantify the suppression of the aging transition in the quantum domain, we solve Eq.~(\ref{eq:3})
directly in the Fock basis for the active and inactive oscillator
populations. The mean-field approximation used in Eq.~(\ref{eq:3}) factorizes the
many-body density matrix into one-body density matrices, with the
interaction between the two populations entering self-consistently
through their mean fields \cite{qaging}.  
The reduced master equations for the active ($s=a$) and inactive
($s=i$) populations can then be written in the unified form
\begin{align}
\dot{\rho}_s={}&
-i[\omega a^\dagger a,\rho_s]+\Gamma_s^-{\cal D}[a]\rho_s+\Gamma_s^+{\cal D}[a^\dagger]\rho_s\nonumber\\
&+k{\cal D}[a^2]\rho_s+V\left({\cal A}_s[a^\dagger,\rho_s]-{\cal A}_s^\ast[a,\rho_s]\right),
\label{eq:two_pop_master}
\end{align}
where, $\Gamma_a^-=\kappa_c$, $\Gamma_a^+=G_a+\kappa_c\gamma$, 
and $\Gamma_i^-=G_b+\kappa_c$, $\Gamma_i^+=\kappa_c\gamma$. 

We expand each one-body density matrix in the truncated Fock basis,
\begin{equation}
\rho_s(t)=\sum_{m,n=0}^{n_c-1}
\rho^{(s)}_{mn}(t)
|m\rangle\langle n|,
\label{eq:Fock_expansion}
\end{equation}
where $n_c$ is the Fock-space cutoff. The equation for an individual matrix element is
\begin{align}
\dot{\rho}^{(s)}_{mn}
={}&
-i\omega(m-n)\rho^{(s)}_{mn}\nonumber\\
&+\Gamma_s^+\left[\sqrt{mn}\,\rho^{(s)}_{m-1,n-1}-\frac{m+n+2}{2}\rho^{(s)}_{mn}\right]\nonumber\\
&+\Gamma_s^-\left[\sqrt{(m+1)(n+1)}\rho^{(s)}_{m+1,n+1}-\frac{m+n}{2}\rho^{(s)}_{mn}\right]\nonumber\\
&+k\left[\sqrt{(m+1)(m+2)(n+1)(n+2)}\rho^{(s)}_{m+2,n+2}\right.\nonumber\\
&\hspace{3.0cm}
\left.
-\frac{m(m-1)+n(n-1)}{2}
\rho^{(s)}_{mn}
\right]\nonumber\\
&+
V{\cal A}_s\left[\sqrt{m}\,\rho^{(s)}_{m-1,n}-\sqrt{n+1}\,
\rho^{(s)}_{m,n+1}\right]\nonumber\\
&-
V{\cal A}_s^\ast\left[\sqrt{m+1}\,
\rho^{(s)}_{m+1,n}
-\sqrt{n}\,
\rho^{(s)}_{m,n-1}
\right].
\label{eq:rho_mn_general}
\end{align}
Matrix elements carrying indices outside the interval
$0\leq m,n\leq n_c-1$ are set to zero. The density matrices are normalized according to ${\rm Tr}\rho_s=\sum_{n=0}^{n_c-1}\rho^{(s)}_{nn}=1$, and also Hermiticity requires $\rho^{(s)}_{mn}=\rho^{(s)\ast}_{nm}$.

Equation~(\ref{eq:rho_mn_general}) constitutes a closed finite-dimensional
nonlinear system of ordinary differential equations because the mean
fields ${\cal A}_a$ and ${\cal A}_i$ are themselves linear functions of
the density-matrix elements. In particular,
\begin{equation}
\alpha_s
= \sum_{n=0}^{n_c-2}\sqrt{n+1}\,
\rho^{(s)}_{n+1,n}.
\label{eq:alpha_fock}
\end{equation}
Consequently, the nonlinear damping parameter $k$ enters the dynamics
both explicitly through Eq.~(\ref{eq:rho_mn_general}) and implicitly
through the dynamically generated Fock-state distribution.

For numerical implementation, Eq.~\ref{eq:3} is represented in a truncated Fock basis, yielding the equivalent finite set of density-matrix-element equations given in Eq.~\ref{eq:rho_mn_general}. We solve these equations self-consistently using QuTiP \cite{qutip}.
Throughout the paper we consider $\omega=2, N=100$ without any loss of generality. We take $ G_j=4 $ for the active elements and $ G_j=2 $ for the inactive elements; the choice is arbitrary for $G_j>0$. To track oscillatory and nonoscillatory (aged) state we use an order parameter based on the mean boson number per oscillator: $ \bar{n}_{mf} = \frac{1}{N} \sum_{j} \langle a_j^\dagger a_j \rangle $: the order parameter is defined as \cite{qaging}
$$
Q = \frac{\bar{n}_{mf}(p)}{\bar{n}_{mf}(0)},
$$
which is the normalized mean boson number, where, $\bar{n}_{mf}(p)$ is the mean boson number per oscillator for a particular $p$ value and $\bar{n}_{mf}(0) $ is the same for $p=0$.

\begin{figure}
 \centering
 \includegraphics[width=.42\textwidth]{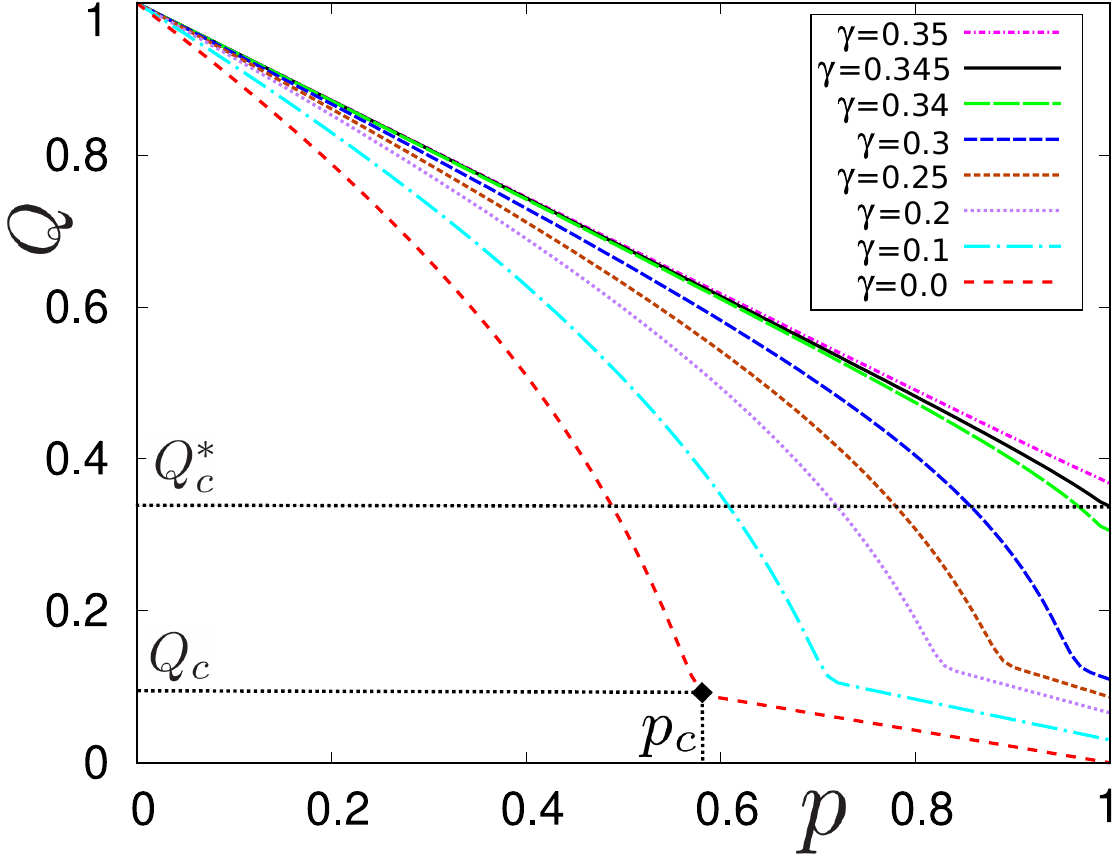}
 \caption{Order Parameter ($ Q $) vs. fraction of inactive oscillators ($p$) for different control parameter $ \gamma $. The `knee' appears at $p=p_c$ gives the onset of the quantum aging transition. Note that, at $\gamma=0.345$ the `knee' reaches $p_c=1$, indicating the disappearance of the aging region. $Q_c$ and $Q_c^*$ are the order parameters at the aging point and revival point, respectively. Other parameters are $ \omega=2, V=5, k=0.2, N=100 $.}
 \label{fig:qaging}
\end{figure}

Before we explore the controlling effect of the single boson pumping, let us discuss in brief the original quantum aging scenario (in the absence of boson pumping term in Eq.~\ref{eq:qm}). For this we take $\gamma=0$ and draw the order parameter $Q$ vs. inactivation ratio, $p$. Figure~\ref{fig:qaging} shows that (for $\gamma=0$) quantum aging is characterized by the rapid  decrease in mean boson number $Q$ with increasing inactivation ratio. Beyond a critical value $p=p_c$, known as the `knee point', the decrement rate becomes drastically less and the variation of $Q$ follows an inclined straight line eventually reaches very close to zero (in the quantum limit) at $p=1$. Ref.~\cite{qaging} designates the `knee point' at $p=p_c$ as the point of quantum aging transition; at the `knee point' $Q=Q_c\ne0$. This is unlike the classical aging transition, where at $p=p_c$, the order parameter $Q$ vanishes \cite{daido}.

\begin{figure}
 \centering
\includegraphics[width=.48\textwidth]{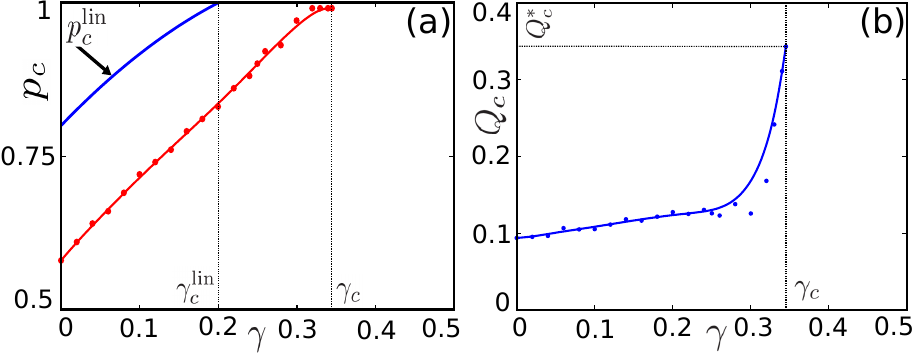}
 \caption{The variation of (a) $p_{c}$ with control parameter $\gamma$. It also shows the variation of $p_c^{\rm lin}$ with $\gamma$ (cf. Eq.~\ref{eq:pc_thermo}), and the corresponding $\gamma_c^{\rm lin}$ (cf. Eq.~\ref{eq:gc_thermo})  (b) The Variation of $Q_c$ with $\gamma$. Other parameters are $ \omega=2, V=5, k=0.2, N=100 $.}
 \label{fig:pcqc}
\end{figure}
Next, we introduce the bosonic pumping term by increasing $\gamma$ from zero: Figure~\ref{fig:qaging} demonstrates the aging transition curves for different $\gamma$. As we increase the value of $\gamma$, two distinct observations can be made: (i) the value of $p_c$ increases, i.e., to achieve aging transition, a large number of nodes need to be in the inactive state. We find that there exists a critical value of $\gamma=\gamma_c$ beyond which no aging transition occurs. In the present case we found $\gamma_c\approx 0.345$: at this value, the `knee' reaches the boundary $p=1$ (let at this point $Q_c=Q_c^*$). For $\gamma>\gamma_c$ the aging transition is completely destroyed and the network remains in the oscillatory state even if all the nodes become inactive (i.e., $p=1$). We illustrate the scenario with an exemplary value $\gamma=0.35~(>\gamma_c)$: in this case, we observe that no `knee' appears and the curve touches the $Q$--axis at $p=1$ with a slope that is almost unchanged with the variation of $p$.  

The above observations are clearly demonstrated with the variations of $p_c$ with $\gamma$ as shown in Fig.~\ref{fig:pcqc}(a). It shows that $p_c$ increases with increasing $\gamma$. As $\gamma$ approaches $\gamma_c$ from below, $p_c$ approaches 1; for $\gamma>\gamma_c$, no aging transition is observed: it indicates that the network can maintain its dynamical robustness even when all the nodes become inactive. The result of the linear stability analysis of the coherent amplitude are also shown in this figure; it shows the variation of $p_c^{\rm lin}$ with $\gamma$ (given by Eq.~\ref{eq:pc_thermo}), and the corresponding $\gamma_c^{\rm lin}$ (given by Eq.~\ref{eq:gc_thermo}). Note that the actual quantum-aging boundary ($p_c$ and $\gamma_c$) obtained from the density-matrix dynamics shifted towards a higher value of $\gamma_c>\gamma_c^{\rm lin}$ and lower value of $p_c<p_c^{\rm lin}$. This mismatch arises due to the following two reasons: First, our linear stability analysis was carried out around the zero-amplitude coherent state, however, in the quantum domain at the aging point ($p=p_c$), the mean boson number $Q=Q_c$ is non zero. Second, the linear results are independent of $k$, however, in reality, nonlinear damping parameter $k$ plays a crucial role in quantum aging and its revival (will be elaborated later in this section). Fig.~\ref{fig:pcqc}(b) depicts the variation of $Q_c$ with $\gamma$. It shows a nonuniform increasing trend with increasing $\gamma$: The rate of increase of $Q_c$ becomes pronounced near $\gamma_c$.
In Fig.~\ref{fig:pcqc}(a,b), $\gamma_c$ is shown by a vertical dashed line to mark the point where the aging state is revoked irrespective of the inactivation ratio: for $\gamma>\gamma_c$ oscillatory state always prevails.

\begin{figure}
 \centering
 \includegraphics[width=0.4\textwidth]{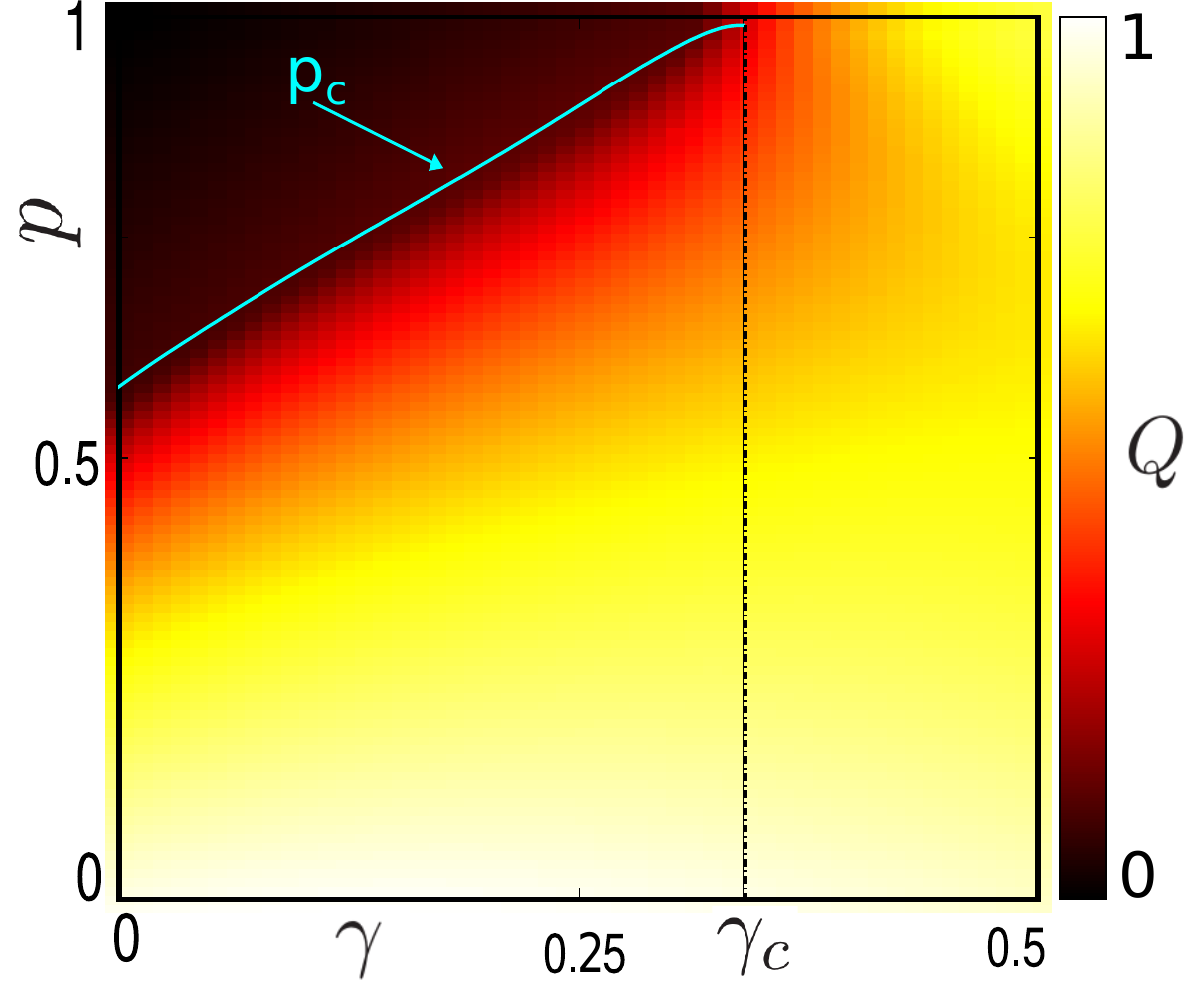}
 \caption{Two parameter plot of $Q$ in $\gamma$-$p$ space. The solid curve in cyan shows the variation of $p_c$ with $\gamma$ above which aging occurs. $\gamma_c=0.345$ is shown by vertical dashed line. Other parameters are $\omega=2, V=5, k=0.2, N=100$.}
 \label{fig:2p}
\end{figure}

To get a global picture of the effect of $\gamma$ on the aging scenario, we examine the variation of the  normalized mean boson number $Q$ in the $\gamma-p$ parameter space, which is shown in Fig.~\ref{fig:2p}. This graph also depicts the variation of $p_c$ beyond which aging occurs (solid curve in cyan). The aging state lies above the $p_c$ curve.  $\gamma_c=0.345$ is shown by vertical dashed line. For $\gamma>\gamma_c$ the normalized mean boson number becomes large even for $p=1$, i.e., when all the nodes become inactive indicating that the network remains in the oscillatory state. 

\begin{figure}
 \centering
 \includegraphics[width=.5\textwidth]{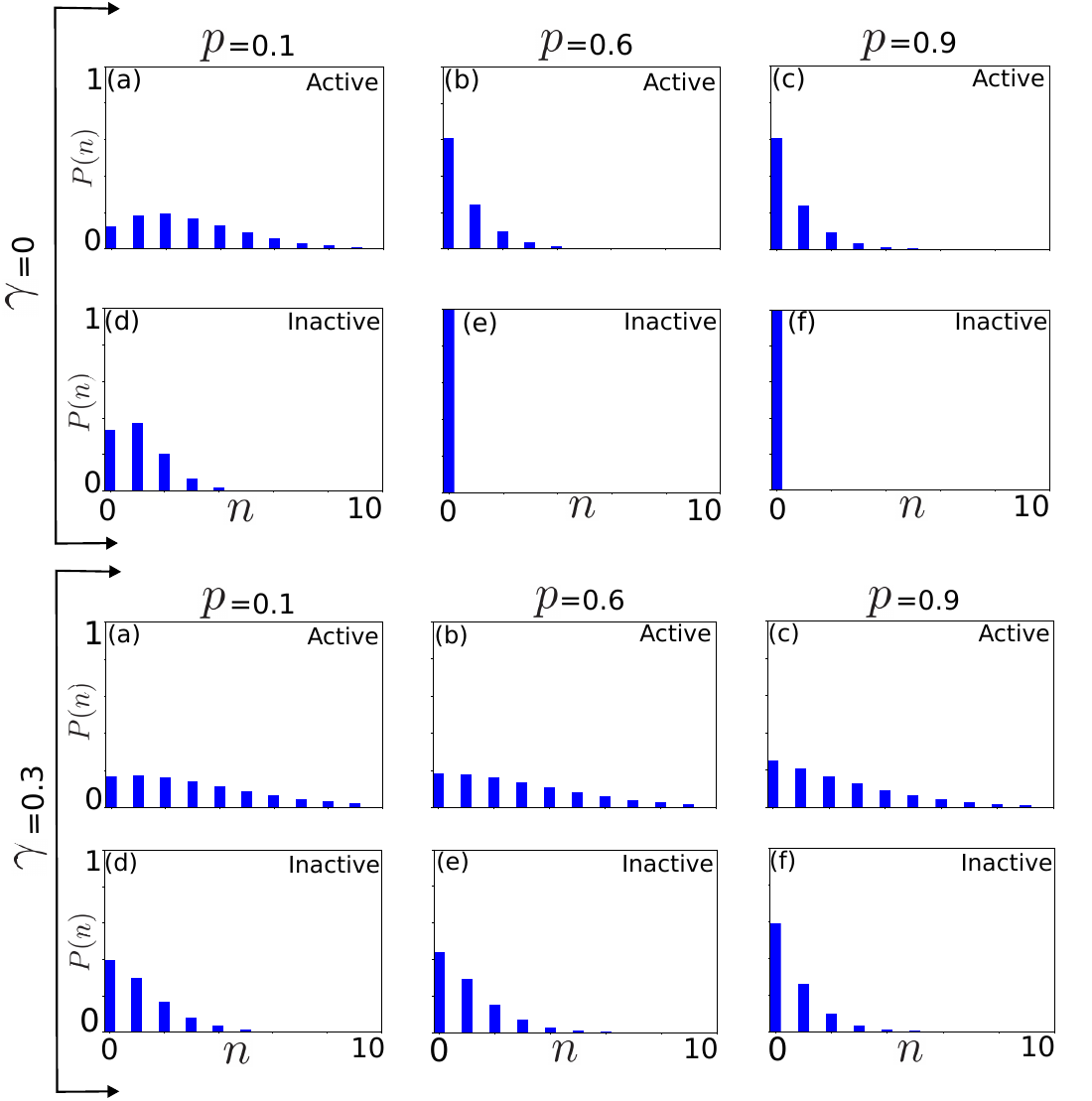}
 \caption{The variation in the boson number distribution $P(n)$ in the Fock space for two different $\gamma$: Upper two panels: $\gamma=0$, Lower two panels: $\gamma=0.3$. Other parameters are $ \omega=2, V=5, k=0.2, N=100 $.}
 \label{fock}
\end{figure}

The mechanism of control of aging and revival of oscillation can be visualized more clearly by tracking the Fock state distribution of the active and inactive nodes (cf. Eq.~\ref{eq:rho_mn_general}, and Eq.~\ref{eq:alpha_fock}).  Figure~\ref{fock} shows the changes in the boson number distribution $P(n)$ in the Fock space. In the network without control (i.e., $\gamma=0$), the  boson number distribution for the active and inactive groups are shown in the upper two panels (for $V=5$): note that at a very low inactivation ratio $p$, the coupling brings the inactive oscillators to higher bosonic state. However, for a larger $p$, when more nodes are inactive, coupling alone can not bring the inactive oscillators to excited states: this can be seen from Fig.~\ref{fock}(e,f) in the upper panel, where inactive nodes occupy only the ground state. However, this scenario is affected appreciably by the control parameter $\gamma$ as shown in the lower two panels of Fig.~\ref{fock}. One observes that even for higher number of inactive oscillators [e.g., $p=0.6$ in (e) and $p=0.9$ in (f) in the lowermost panel], an increasing $\gamma$ is able to make the inactive nodes to populate the excited states thereby promoting rhythmicity.

\begin{figure}[t]
 \centering
 \includegraphics[width=.5\textwidth]{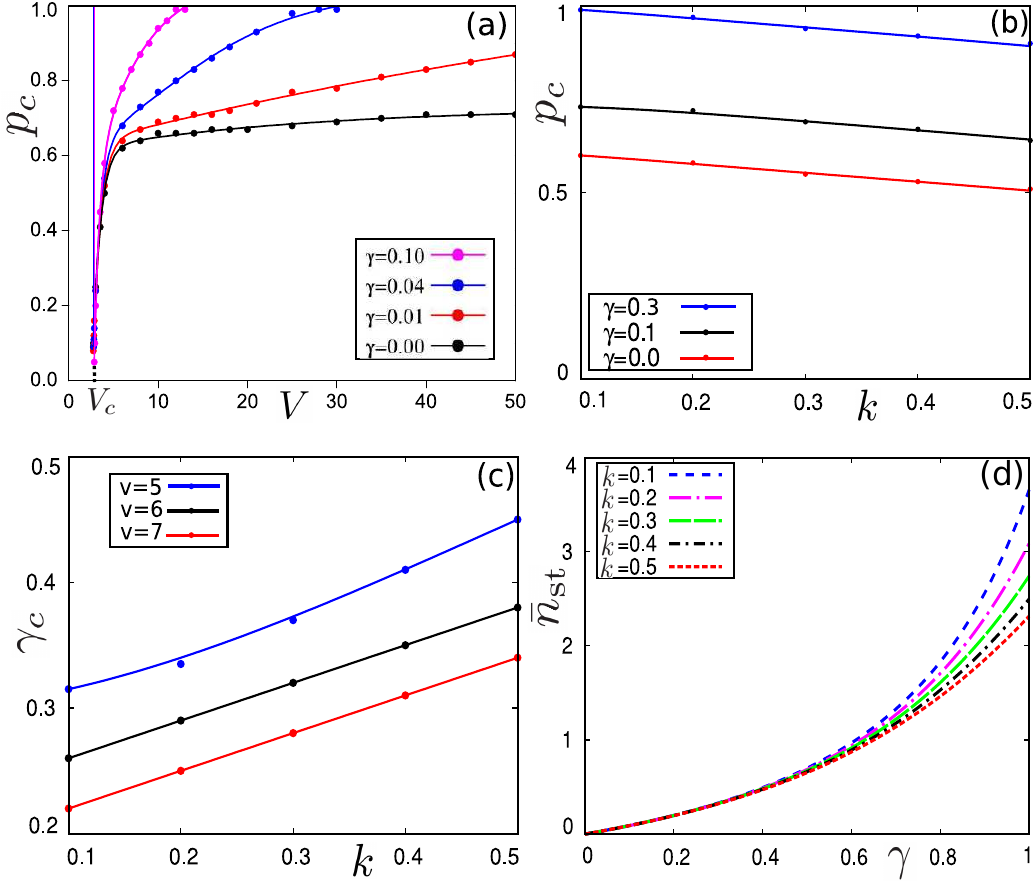}
 \caption{(a) Plot of $p_c$ versus $V$ for different $\gamma$ ($k=0.2$). (b) Plot of $p_c$ versus $k$ for different $\gamma$ ($V=5$). (c) The variation of $\gamma_c$ with $k$ ($V=5$). (d) Variation of ${n}_{\rm st}$ with $\gamma$ (Eq.\ref{eq:nst_Vkgamma}) for different $k$ ($V=5$). Other parameters are $\omega=2$, $N=100$.}
 \label{fig:vk}
\end{figure}

Next, we explore the effect of the control parameter $\gamma$ on the revival mechanism in the $V-p_c$ space (shown in Fig.~\ref{fig:vk}(a)). We observe that, there exists a lower value of coupling strength $V_c$, below which no aging is possible. Interestingly, the critical value $V_c$ is almost independent of $\gamma$, which is indicated by the vertical line at $V_c$ for different $\gamma$ value. It can be seen that the area of the aging state in the $V-p_c$ parameter space (area made by the curve and the vertical line with the top axis at $p_c=1$) shrinks appreciably with even a slight increase in $\gamma$.  For $\gamma\ge\gamma_c$, the area becomes zero (not shown) indicating that no aging takes place and the rhythmicity is established in the whole network.

We further found that the revival process from quantum aging is affected by the nonlinear damping parameter $k$, which is in stark contrast to classical case where the revival process does not depend upon the nonlinear damping parameter $k$ (cf. Eq.\ref{eq:gc_thermo}) \cite{liu2016enhancing}. 
Figure~\ref{fig:vk}(b) shows the effect of nonlinear damping on $p_c$ in the quantum system for different $\gamma$. One can observe that a larger $k$ results in a lesser $p_c$ meaning a lesser fraction of inactive nodes lead to the aging transition. 
Finally, in Fig.~\ref{fig:vk}(c) we show that with increasing $k$ the required $\gamma_c$ for the revival of oscillation also increases: this can be explained from the fact that a larger $k$ strengthens two-boson loss and suppresses the population of higher Fock states. Consequently, stronger incoherent pumping is required to compensate this depletion and eliminate the aging transition, leading to an increase of the critical pumping strength $\gamma_c$ with $k$.

The microscopic origin of the revival can be understood physically from the competition between the single-boson loss, controlled single-boson pumping, and nonlinear two-boson damping processes in Eq.~\ref{eq:3}. Let us first consider the maximally degraded limit, $p=1$, for which all oscillators are
inactive and $G_j=G_b$; therefore, for this state,
$\langle a\rangle=0$, the mean-field contribution vanishes and Eq.~\ref{eq:two_pop_master}
reduces to
\begin{equation}
\dot{\rho}
=
-i[\omega a^\dagger a,\rho]
+\Gamma^-{\cal D}[a]\rho
+k{\cal D}[a^2]\rho
+\Gamma_p{\cal D}[a^\dagger]\rho .
\label{eq:D1}
\end{equation}
where $\Gamma^- =G_b+\kappa_c$, and as before $\Gamma_p=\kappa_c\gamma$. 
To elucidate the resulting population dynamics, we expand the stationary density matrix in the Fock basis,
$ \rho^{(0)} =\sum_{n=0}^{\infty}P_n |n\rangle\langle n|$, $\sum_{n=0}^{\infty}P_n=1$. Substitution into Eq.~(\ref{eq:D1}) gives
\begin{align}
\dot{P}(n)={}&\Gamma_p\left[nP(n-1)-(n+1)P(n)\right]\nonumber\\
&+\Gamma^-\left[(n+1)P(n+1)-nP(n)\right]\nonumber\\
&+k\left[(n+1)(n+2)P(n+2)-n(n-1)P(n)\right].
\label{eq:D4}
\end{align}
Equivalently, \eqref{eq:D4} gives the time evolution of the mean boson number as
\begin{equation}
\frac{d\bar n}{dt}=\Gamma_p(\bar n+1)-\Gamma^-\bar n-2k\langle n(n-1)\rangle .
\label{eq:17}
\end{equation}
The first two terms represent incoherent boson injection (controlled through $\gamma$) and linear
loss, respectively, while the last term describes nonlinear two-boson loss governed by the nonlinear damping parameter $k$.
Under the approximation $\langle n(n-1)\rangle\simeq\bar n^2$, in the thermodynamic limit ($\kappa_c\approx2V$), one gets the stationary mean boson number as 
\begin{equation}
\bar{n}_{\rm st}\simeq\frac{\sqrt{\left[G_b+2V(1-\gamma)\right]^2+16kV\gamma}-\left[G_b+2V(1-\gamma)\right]}{4k}.
\label{eq:nst_Vkgamma}
\end{equation}
Equation~(\ref{eq:nst_Vkgamma}) shows that increasing $\gamma$ increases the stationary excitation level, which also depends upon the nonlinear damping $k$: Fig.~\ref{fig:vk} illustrates this for different $k$.
By restoring the population of excited Fock states, the pumping supplies the excitation required for the inactive oscillators to respond to the mean field and participate in the collective dynamics, thereby eliminating the quantum aging transition and restoring rhythmicity.



\section{Conclusions}
\label{sec:con}
In this work, we have demonstrated the revival of rhythmic dynamics from
an aged state in a network of coupled quantum self-sustained oscillators
through controlled incoherent single-boson pumping. By introducing a
tunable pumping channel in the Lindblad dynamics, we find that the aging
transition can be progressively suppressed as the pumping strength is
increased, and can eventually be completely revoked. In this regime, the
network retains rhythmic dynamics even when all oscillators become
inactive, demonstrating a substantial enhancement of its dynamical
robustness. 
The Fock-state redistribution in the inactive oscillators explains the revival: incoherent pumping restores higher-occupation states, while nonlinear damping limits their growth and influences the aging threshold.

A central finding of our study is that complete compensation of the
dissipative effect associated with the global coupling is not necessary
for the revival of rhythmicity. The aging transition is eliminated at a
pumping strength substantially below the full-compensation limit. This
shows that relatively weak incoherent pumping can be sufficient to
restore collective dynamical activity in a severely degraded quantum
network.

These results establish controlled incoherent single-boson pumping as a mechanism for revoking quantum aging and restoring rhythmicity. With the recent progress in quantum engineering \cite{science-expt-vdp,expt-sync-qvdp}, we believe that the proposed mechanism is experimentally plausible in platforms such as cavity optomechanical and trapped-ion systems, where sideband engineering provides controlled gain and loss of bosonic excitations. In cavity optomechanical systems, controlled incoherent boson pumping can be implemented through blue-detuned driving on the single-sideband transition \cite{expt-opto,brud_prl1}. A single-boson pumping process can also be realized in trapped-ion systems \cite{lee_prl} as well.

More broadly, our results show that environmental engineering can be used not only to compensate dissipation but also to restore collective dynamical functionality in quantum networks. This provides a route toward more robust quantum networks operating in the presence of unavoidable degradation and loss and thus the present study can be extended to the other quantum network, e.g. a network of qubits \cite{qubit-aging,qbit-aging-2}. We believe that our results are beneficial to the understanding and development of a more robust network in the context of quantum technology.

\section{acknowledgments}
B.P. acknowledges the financial assistance from DST-INSPIRE, India. 

\section*{Data availability}
The datasets generated during and/or analyzed during the current study are available from the corresponding author on reasonable request.

\appendix
\begin{widetext}
\section{Derivation of Eq.~\ref{eq:3}}\label{app:a}
For derivation of Eq.~\ref{eq:3} let us consider $N=3$, and then generalize it for any $N$. For $N=3$ the master equation (\ref{eq:qm}) reduces to
\begin{multline}
\begin{split}
\label{eq:quan_mas_red}
\dot \rho &= \sum_{j=1}^{3} \Bigl(-i[ \omega a_j^\dagger a_j,\rho] + G_j \mathcal{D} [O_j](\rho) + k \mathcal{D} [{a_j}^2] (\rho)\Bigl) + \frac{V}{N} \sum_{j=1}^{3} \sideset{}{'} \sum_{{j^\prime}=1}^{3} \mathcal{D} [a_j - a_{j^\prime}](\rho)+ \frac{2V(N-1) \gamma}{N} \sum_{j=1}^{3} \mathcal{D} [a_j^\dagger] (\rho)
\end{split}
\end{multline}
We will show the contribution of the dissipator containing the global diffusive coupling only. The contributions of rest of the terms are straight forward. The coupling term in an expanded form reads
\begin{multline}
\begin{split}
\label{eq:diss_all}
\frac{V}{N} \sum_{j=1}^{3} \sideset{}{'} \sum_{{j^\prime}=1}^{3} \mathcal{D} [a_j - a_{j^\prime}](\rho)&=  \frac{V}{N} \Bigl(\mathcal{D} [a_1 - a_2](\rho) + \mathcal{D} [a_1 - a_3](\rho) + \mathcal{D} [a_2 - a_1](\rho) \\
&+\mathcal{D} [a_2 - a_3](\rho) +\mathcal{D} [a_3 - a_1](\rho) +\mathcal{D} [a_3 - a_2](\rho)\Bigl).
\end{split}
\end{multline}
We have to evaluate all the dissipator terms: here we show only the derivation of the first dissipator; rest will follow the same technique. Using the relation $\mathcal{D} [\hat L](\hat \rho) = \hat L \hat \rho \hat L^\dagger - \frac{1}{2} \{\hat L^\dagger \hat L ,\hat \rho\}$, we get
\begin{equation}
\begin{split}
\label{eq:diss11}
\mathcal{D} [a_1 - a_2](\rho) &= (a_1-a_2)\rho(a_1-a_2)^\dagger - \frac{1}{2}\Bigl((a_1-a_2)^\dagger(a_1-a_2)\rho + \rho(a_1-a_2)^\dagger(a_1-a_2)\Bigl) \\
&= a_1\rho {a_1}^\dagger - a_1\rho {a_2}^\dagger - a_2\rho a_1^\dagger + a_2 \rho a_2^\dagger - \frac{1}{2}\bigl(a_1^\dagger a_1\rho - a_1^\dagger a_2\rho \\
&- a_2^\dagger a_1\rho + a_2^\dagger a_2 \rho + \rho a_1^\dagger a_1 - \rho a_1^\dagger a_2 - \rho a_2^\dagger a_1 + \rho a_2^\dagger a_2\bigl).
\end{split}
\end{equation}
All these identities can be evaluated using 
\begin{equation}
\begin{split}
\label{eq:tensor}
a_2\rho a_1^\dagger &= (\mathbf{1} \otimes a_2 \otimes \mathbf{1})(\rho_1 \otimes \rho_2 \otimes \rho_3)(a_1^\dagger \otimes \mathbf{1} \otimes \mathbf{1}) \\
&= \rho_1 a_1^\dagger \otimes a_2 \rho_2 \otimes \rho_3.
\end{split}
\end{equation}
Regathering all the terms, we get $\mathcal{D} [a_1 - a_2](\rho)$ as follows,
\begin{equation}
\begin{split}
\label{eq:diss12}
\mathcal{D} [a_1 - a_2](\rho) &= a_1\rho_1 a_1^\dagger \otimes \rho_2 \otimes \rho_3 - a_1 \rho_1 \otimes \rho_2 a_2^\dagger \otimes \rho_3 - \rho_1 a_1^\dagger \otimes a_2 \rho_2 \otimes \rho_3 + \rho_1 \otimes a_2 \rho_2 a_2^\dagger \otimes \rho_3 \\
&- \frac{1}{2}\bigl(a_1^\dagger a_1 \rho_1 \otimes \rho_2 \otimes \rho_3 - a_1^\dagger \rho_1 \otimes a_2 \rho_2 \otimes \rho_3 - a_1 \rho_1 \otimes a_2^\dagger \rho_2 \otimes \rho_3 + \rho_1 \otimes a_2^\dagger a_2 \rho_2 \otimes \rho_3 \\
&+ \rho_1 a_1^\dagger a_1 \otimes \rho_2 \otimes \rho_3 - \rho_1 a_1^\dagger \otimes \rho_2 a_2 \otimes \rho_3 - \rho_1 a_1 \otimes \rho_2 a_2^\dagger \otimes \rho_3 + \rho_1 \otimes \rho_2 a_2^\dagger a_2 \otimes \rho_3\bigl).
\end{split}
\end{equation}
In order to calculate the contribution of ${\mathcal{D} [a_1 - a_2](\rho)}$ towards the density matrix of first oscillator ($\rho_1$) we need to use the notion of partial trace as given by: 
\begin{equation}
\begin{split}
\label{eq:diss13}
Tr_3(Tr_2(\mathcal{D} [a_1 - a_2](\rho)))&= a_1 \rho_1 a_1^\dagger - a_1 \rho_1 \langle a_2^\dagger \rangle - \rho_1 a_1^\dagger \langle a_2 \rangle + \rho_1 \langle a_2^\dagger a_2 \rangle - \frac{1}{2}\bigl(a_1^\dagger a_1 \rho_1 - a_1^\dagger \rho_1 \langle a_2 \rangle \\
&- a_1 \rho_1 \langle a_2^\dagger \rangle + \rho_1 \langle a_2^\dagger a_2 \rangle + \rho_1 a_1^\dagger a_1 - \rho_1 a_1^\dagger \langle a_2 \rangle - \rho_1 a_1 \langle a_2^\dagger \rangle + \rho_1 \langle a_2^\dagger a_2 \rangle\bigl). \\
&= \mathcal{D} [a_1](\rho_1) + \frac{1}{2} [a_1^\dagger, \rho_1] \langle a_2 \rangle - \frac{1}{2} [a_1, \rho_1] \langle a_2^\dagger \rangle.
\end{split}
\end{equation}
In a similar way, we calculate the contribution of all the dissipator terms in \eqref{eq:diss_all} in determining the density matrix of the first oscillator and adding them we get 
\begin{equation}
\begin{split}
\label{eq:com1}
\dot \rho_{1^{\prime}} = \frac{2V.(3-1)}{3} \mathcal{D} [a_1](\rho_1) + \frac{V}{3}\left(a_1^\dagger, \rho_1] \langle a_2 \rangle - [a_1, \rho_1] \langle a_2^\dagger \rangle + [a_1^\dagger, \rho_1] \langle a_3 \rangle - [a_1, \rho_1] \langle a_3^\dagger \rangle \right)
\end{split}
\end{equation}
Generalizing for the $N$ oscillators here, the resulting \eqref{eq:com1} gives the contribution towards the fourth and fifth term of Eq.~\ref{eq:3}
\begin{equation}
\begin{split}
\label{eq:com2}
\dot \rho_{j^{\prime}} &= \frac{2V(N-1)}{N} \mathcal{D} [a_j](\rho_j) + \frac{V}{N}\left([a_j^\dagger, \rho_j] \sideset{}{'} \sum_{{j^\prime}=1}^{N} {\langle a_{j^\prime} \rangle}_j - [a_j, \rho_j] \sideset{}{'} \sum_{{j^\prime}=1}^{N} {\langle a_{j^\prime}^\dagger \rangle}_j\right) \\
&= \frac{2V(N-1)}{N} \mathcal{D} [a_j](\rho_j) + V \left( \mathcal{A}[a_j^\dagger,\rho_j] - {\mathcal{A}^*} [a_j, \rho_j]\right),
\end{split}
\end{equation}
where $\mathcal{A}$ and ${\mathcal{A}^*}$ are given by: $\mathcal{A} = \frac{1}{N}\sideset{}{'} \sum_{{j^\prime}=1}^{N} {\langle a_{j^\prime} \rangle}_j$ and ${\mathcal{A}^*} = \frac{1}{N}\sideset{}{'} \sum_{{j^\prime}=1}^{N} {\langle a_{j^\prime}^\dagger \rangle}_j$. 
\end{widetext}

\providecommand{\noopsort}[1]{}\providecommand{\singleletter}[1]{#1}%
\end{document}